\documentclass[aps, prb, reprint, superscriptaddress]{revtex4-1}
\usepackage{amsmath, amssymb, times, mathrsfs, hyperref, array, bbm}
\usepackage{graphicx}
\usepackage[usenames,dvipsnames]{xcolor}
\usepackage{braket}

\hypersetup{colorlinks=false}

\bibpunct{[}{]}{,}{n}{}{}

\renewcommand{\vec}[1]{{\boldsymbol #1}}
\usepackage{graphicx}

\definecolor{green(html/cssgreen)}{rgb}{0.0, 0.5, 0.0}

\begin{document}
\title{Metallization and proximity superconductivity in topological insulator nanowires}

\author{Henry F. Legg}
\affiliation{Department of Physics, University of Basel, Klingelbergstrasse 82, CH-4056 Basel, Switzerland}

\author{Daniel Loss}
\affiliation{Department of Physics, University of Basel, Klingelbergstrasse 82, CH-4056 Basel, Switzerland}

\author{Jelena Klinovaja}
\affiliation{Department of Physics, University of Basel, Klingelbergstrasse 82, CH-4056 Basel, Switzerland}

%\date{\today}

\begin{abstract}

A heterostructure consisting of a topological insulator (TI) nanowire brought into proximity with a superconducting layer provides a promising route to achieve topological superconductivity and associated Majorana bound states (MBSs). Here, we study effects caused by such a  coupling between a thin layer of an $s$-wave superconductor and a TI nanowire. We show that there is a distinct phenomenology arising from the metallization of states in the TI nanowire by the superconductor. In the strong coupling limit, required to induce a large superconducting pairing potential, we find that metallization results in a shift of the TI nanowire subbands ($\sim 20$~meV) as well as it leads to a small reduction in the size of the subband gap opened  by a magnetic field applied parallel to the nanowire axis. Surprisingly, we find that metallization effects in TI nanowires can also be beneficial. Most notably, coupling to the superconductor induces a potential in the portion of the TI nanowire close to the interface with the superconductor, this breaks inversion symmetry and at finite momentum lifts the spin degeneracy of states within a subband. As such coupling to a superconductor can create or enhance the subband splitting that is key to achieving topological superconductivity. This is in stark contrast to semiconductors, where it has been shown that metallization effects always reduce the equivalent subband-splitting caused by spin-orbit coupling. We also find that in certain geometries metallization effects can reduce the critical magnetic required to enter the topological phase. We conclude that, unlike in semiconductors, the metallization effects that occur in TI nanowires can be relatively easily mitigated, for instance by modifying the geometry of the attached superconductor or by compensation of the TI material.
\end{abstract}

\maketitle

%==============================================================================================
%    Introduction
%==============================================================================================

\section{Introduction}

Topological superconductors can host localized excitations known as Majorana fermions, for instance at their boundaries or within vortices \cite{volovik1999,read2000}. It has been argued that the non-Abelian statistics and localized nature of Majorana fermions make them promising candidates for quantum computation \cite{kitaev2001,Ivanov2001,kitaev2003}. One route to achieve topological superconductivity is via the proximity effect in which a trivial $s$-wave superconductor induces a pairing potential in another material. For instance, due to the proximity effect a superconductor coupled to the surface of a topological insulator (TI) can give rise to a $p_x+i p_y$ superconductivity where Majorana bound states (MBSs) are predicted to appear at the center of vortices \cite{Fu2008,hasan2010,sato2017}.

Another proposed platform to achieve MBSs is a thin nanowire of three-dimensional topological insulators with quantum confined surface states \cite{Breunig2021}. When brought into proximity with a superconductor, as it has been shown theoretically, a topological phase of large extend in the parameter space can be achieved in a TI nanowire, either when the superconducting order parameter possesses a vortex \cite{Cook2011} -- essentially requiring a full superconducting shell \cite{deJuan2019} --  or without a vortex \cite{Legg2021} -- requiring the breaking of inversion symmetry by, for example, the application of a gate. Since a full superconducting shell is not experimentally feasible in the near term, we will focus on  setups without a vortex in the order parameter.

In the past few years there has been substantial progress in the growth of thin TI nanowire devices \cite{Breunig2021,Peng2010,hong2012,hamdou2013,cho2015,jauregui2015,jauregui2016,Ziegler2018,rosenbach2020,rosenbach2021}, including the growth of bulk insulating devices in which the chemical potential can be tuned close to the Dirac point \cite{Munning2021,legg2021MCA}.  In TI nanowires several transport signatures of the quantum confinement of surface states have been reported, previously these consisted of conductivity oscillations \cite{vishwanath2010,Bardarson2010} as a function of magnetic field or gate voltage \cite{Peng2010,hong2012,hamdou2013,cho2015,jauregui2015,jauregui2016,Ziegler2018,rosenbach2020,rosenbach2021,Munning2021}. Recently, a non-reciprocal transport effect in bulk insulating nanowires provided strong evidence not only for quantum confinement of the surface states but also for the splitting of initially spin-degenerate subbands that is predicted due to the breaking of inversion symmetry by a non-uniform potential through the cross-section of the nanowire \cite{legg2021MCA}. This subband-splitting is a key element required to achieve MBSs without a vortex \cite{Legg2021}.

\begin{figure}[t]
	\centering
  \includegraphics[width=1\columnwidth]{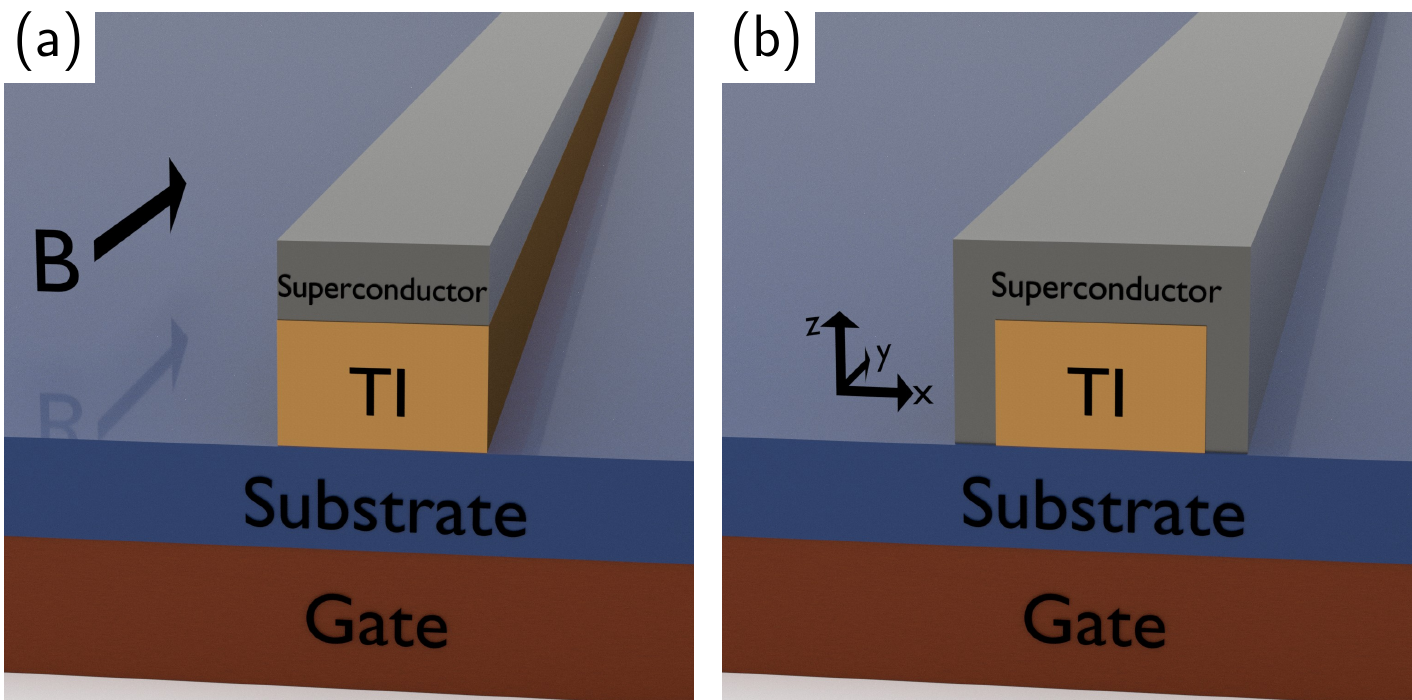}
	\caption{{\bf Heterostructure to generate Majorana bound states in TI nanowires:} A TI nanowire (gold color) is brought into proximity with a superconductor (gray color). The chemical potential in the TI nanowire is adjusted by means of a gate (red color) placed below the nanowire. Additional gates (not shown) can be added to provide further control over the uniformity of the potential in the nanowire. We consider two setups: {\bf (a)} the superconductor is placed only on the top of the nanowire and {\bf (b)} the superconductor has a horseshoe shape and is in contact with three sides of the nanowire.}
\label{fig:setup}
\end{figure}

Given the substantial recent progress in TI nanowire fabrication, the experimental focus now turns to devices where superconductivity is induced in the TI nanowire via the proximity effect \cite{Breunig2021,chen2018,jauregui2018,Schuffelgen2019,kayyalha2019,bai2020,rosenbach2021sc,bai2021,fischer2021}. Such a device is shown in Fig.~\ref{fig:setup} where the TI nanowire is strongly tunnel coupled to a thin layer ($\sim 10$~nm) of superconductor. In semiconductors that possess strong Rashba spin-orbit coupling -- which have also been predicted to host MBSs in certain circumstances \cite{prada2020,laubscher2021} -- the full influence of the tunnel coupling to a superconducting layer has previously been analysed \cite{reeg2017a,reeg2017,reeg2018,antipov2018,woods2018,mikkelsen2018} and revealed a significant roadblock to achieving MBSs in semiconductors is the metallization of the semiconductor by the superconductor. Metallization \cite{reeg2017,reeg2018} is the process by which, when a large pairing potential is induced in the semiconductor, other properties of the semiconductor are renormalized towards the values of the attached superconducting metal. Namely, it was shown in Ref.~\citenum{reeg2018} that inducing a pairing potential of the order found in the parent superconductor ($\Delta \sim \Delta_0$) -- referred to as \emph{strong coupling} -- also results in a large subband-shift ($\sim 100$~meV), a significant increase in the effective mass ($m^*\sim m_e$), and a considerable reduction of the $g$-factor and the spin-orbit energy. In other words, a large proximity induced gap also results in a considerable reduction of the material properties that made the semiconductor desirable for achieving MBSs in the first place.

In this paper we consider the impact of tunnel coupling a thin layer of superconductor to a TI nanowire. Since a TI is closer to a metallic state than a semiconductor and the two-dimensional cross-section of a TI nanowire is fundamental to its nature, we can expect a unique impact of the superconductor on the TI nanowire. Indeed, whilst we find that TI nanowires undergo some similar metallization effects to semiconductors in the strong coupling limit, the size of the metallization effects are smaller than in semiconductors, for instance we find a subband-shift $\sim 20$~meV and a small reduction in the gap induced by orbital effects from a magnetic field parallel to the TI nanowire. Unlike in semiconductors, we find that metallization effects in TI nanowires can actually be beneficial. For instance, in a semiconductor the spin-orbit energy is reduced by the coupling to a superconductor, however, in a TI nanowire the subband splitting energy can actually be created or enhanced in the strong coupling regime since the superconductor provides an inversion symmetry breaking potential similar to that produced by a gate. We also show that  the superconducting gap is highly momentum dependent for certain device geometries such that the magnetic field strength required to reach the topological phase can be reduced due to metallization effects. Our findings show that, unlike in semiconductors, metallization effects in TI nanowires can relatively easily be mitigated and do not pose as significant a roadblock to achieving MBSs.

The paper is organized as follows. In Sec.~\ref{sec:tsc} we discuss the main ingredients that are required to achieve topological superconductivity in TI nanowires, as outlined in Ref.~\citenum{Legg2021}. In Sec~\ref{sec:model} we present the tight binding model that we use to investigate the coupling of a superconducting layer to a TI nanowire. Sec.~\ref{sec:metal} investigates the impact metallization effects have on the features of a TI nanowire that are required to achieve MBSs. This is done first without and then including the effects of a non-uniform potential induced by a gate. Sec.~\ref{sec:exp} outlines how the metallization in TI nanowires can be experimentally mitigated and also compares these metallization effects to the equivalent effects that occur in semiconductor devices. Finally in Sec.~\ref{sec:dis} we discuss the wider implications of our findings for the prospects of realizing MBSs in TI nanowire devices.

\label{sec:intro}

\section{Topological superconductivity\\
in TI nanowires}\label{sec:tsc}

\begin{figure}[t]
	\centering
  \includegraphics[width=1\columnwidth]{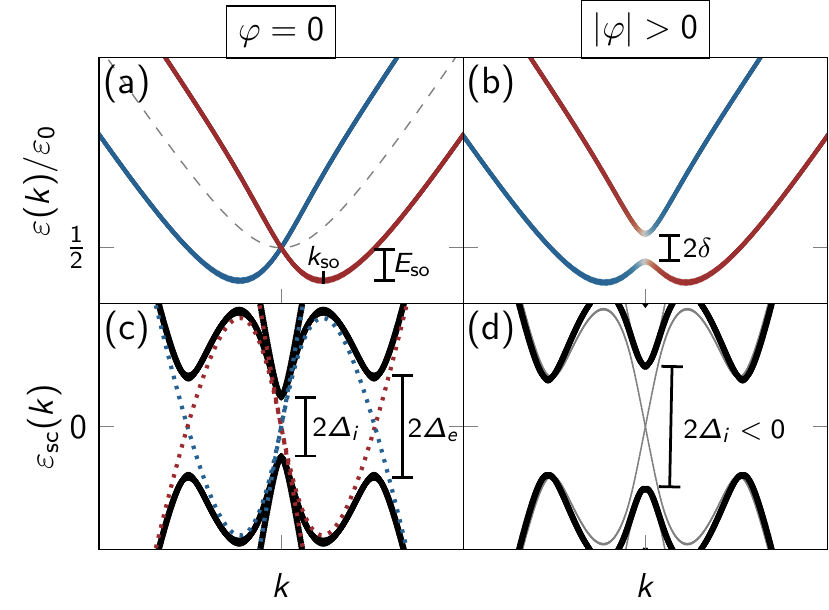}
	\caption{{
\bf Topological superconductivity in TI nanowires:} {\bf (a)} The low-energy spectrum of the TI nanowire consists of doubly degenerate subbands with opposite angular momenta  $\ell=\pm\frac{1}{2},\pm \frac{3}{2},\dots$ (gray dashed line), these are equally spaced in energy by $\epsilon_0=2\pi\hbar v/P$. Applying a gate voltage through the nanowire cross-section lifts the inversion symmetry and thereby
the degeneracy of the subbands at finite momentum, inducing a net spin polarisation (indicated by red/blue color). {\bf (b)} Due to primarily orbital effects, a magnetic field, $B$, applied parallel to the nanowire opens a gap of size $2\delta$ at $k=0$ leaving only a pair of (almost) helical modes at the Fermi-level when the chemical potential is within this gap. {\bf (c)} Setting the chemical potential to the subband crossing at $k=0$. If, in addition to the gating effects but in absence of a magnetic field, a pairing potential $\Delta$ is induced in the TI -- e.g. by tunnel coupling to a superconducting layer -- a superconducting gap opens. In general this will be dependent on the momentum $k$ with an interior gap $\Delta_i$ at $k=0$ and exterior gap $\Delta_e$ at finite momentum. {\bf (d)} As the strength of the parallel magnetic field is increased, the interior gap $\Delta_i$ closes (indicated by the gray line) and then reopens with opposite sign, indicating the transition to a topological superconducting phase and within this phase well-localized MBSs can appear at the ends of the TI nanowire \cite{Legg2021}.}
	\label{fig:schematic}
\end{figure}

We begin by outlining the main ingredients that are required to achieve topological superconductivity and Majorana bound states in TI nanowires, further details can be found in Ref.~\citenum{Legg2021}. The heterostructure we consider is shown in Fig.~\ref{fig:setup}. It consists of a TI nanowire (gold color) that is tunnel coupled to a superconductor (gray color), the chemical potential in the TI nanowire can be adjusted via a gate below the nanowire (red color). The superconductor can either be attached only to the top of the nanowire (Fig.~\ref{fig:setup}a) or to multiple surfaces such as in a horseshoe arrangement (Fig.~\ref{fig:setup}b). MBSs at the ends of the nanowire can be generated in this setup when a magnetic field $\vec B$ applied parallel to the nanowire axis results in a topological superconducting phase (see Fig.~\ref{fig:schematic}).

We first consider the TI nanowire when there is no tunnel coupling to the superconductor. If, in addition, no gate voltage is applied, the spectrum of the TI nanowire is given by \cite{Cook2011,Legg2021}
\begin{equation}
\epsilon_\ell(k)=\pm \hbar v \sqrt{k^2+(2\pi \ell/P)^2},\label{eq:TIspec}
\end{equation}
where $\ell=\pm\frac{1}{2},\pm \frac{3}{2},\dots$, $v$ is the Fermi velocity, $k$ is the momentum along the wire, and  and $P$ is the perimeter of the nanowire [see dashed gray line in Fig.~\ref{fig:schematic}(a)]. The spectrum of Eq.~\eqref{eq:TIspec} describes doubly degenerate subbands which are equally spaced by $\epsilon_0=2\pi \hbar v/P$ as has been shown to hold for many different TI nanowire cross-sections \cite{deJuan2019}. For instance, the left panel of Fig.~\ref{fig:bands} shows the spectrum for a rectangular cross-section without coupling to the superconductor (see next section for details). Throughout this study we will consider a rectangular cross-section -- such a cross-section is produced, for example, by etching MBE films \cite{legg2021MCA}-- although the metallization phenomena we find will be generally applicable to any nanowire cross-section. 

Applying a gate voltage, one induces a  spatially non-uniform potential in the nanowire cross-section \cite{Legg2021}, $\mu(\theta)=\mu_0+\delta\mu(\theta)$, where $\mu_0$ is the average of the potential and $\delta\mu(\theta)=2\sum_{n=1}^\infty \mu_n \cos n\theta$ is the non-uniform component, with $\theta$  the polar angle within the nanowire cross-section. The non-uniform component arises because the gate is located only on one side of the nanowire. This non-uniform potential breaks inversion symmetry and lifts the degeneracy of the subbands at finite momentum $k$ along the nanowire  [see Fig.~\ref{fig:schematic}(a)]. For a non-uniformity with components $\mu_n$ that are small compared to the subband spacing the spectrum becomes~\cite{Legg2021}
\begin{equation}
\varepsilon_{\ell}^\pm(k)\approx \epsilon_\ell(k)\pm\frac{\mu_{2\ell}k}{\sqrt{k^2+(2\pi\ell/P)^2}}+\mu_0,\label{eq:TIspecgated}
\end{equation}
where we have used the original angular momenta $\ell>0$ to label the split subband pair. As such the presence of the non-uniform chemical potential results in a subband-splitting with an energy scale $E_{\rm so}\approx \mu_{2\ell}^2/(2\ell v/R)$ [see Fig.~\ref{fig:schematic}(a)]. Note that the non-uniform component of the chemical potential $\delta\mu(\theta)$ does not alter the energy of the states at $k=0$ and they are only shifted by the average of the gate induced potential $\mu_0$, this effect is associated with Klein-tunnelling.
 
A magnetic field, $B$, applied parallel to the nanowire -- indicated by the black arrow in Fig.~\ref{fig:setup} -- opens a gap of size $\delta\approx h v |\varphi|/P$ at $k = 0$ primarily due to orbital effects, where $\varphi=\Phi/\Phi_0$ is the magnetic flux $\Phi=BA_\perp$ (here  $A_\perp$ is the cross-sectional area of the nanowire) in units of the fundamental flux quantum $\Phi_0= h/e$. If the chemical potential lies within this magnetic field induced gap it leaves only a pair of (almost) helical modes \cite{streda,CNTJK,RibbonJK} at finite Fermi momenta [see Fig.~\ref{fig:schematic}(b)]. 

When the TI nanowire is brought into proximity with an $s$-wave superconductor, a pairing potential $\Delta(\varphi)$ is induced in the TI. It should be noted that in general the induced pairing potential is momentum dependent (see Fig.~\ref{fig:schematic}). In particular, as a result of a non-uniform potential, the wavefunctions of states close to $k=0$ are more uniformly distributed around the nanowire perimeter than states at large momenta (see below). As a result of these differently localized wavefunctions we can identify two gaps in the spectrum of the resulting energy spectrum: An interior gap $\Delta_i$ at $k=0$ and an exterior gap $\Delta_e$ at large momenta (see Fig.~\ref{fig:schematic}c). If in addition to the proximity induced pairing, a magnetic field is applied along the nanowire, then a topological superconducting phase can be achieved when \cite{Legg2021}
\begin{equation}
\delta^2>\Delta^2_{\rm i}(\varphi=0)+\mu^2,\label{eq:topcrit}
\end{equation} where $\mu$ is measured from the subband crossing point at $k=0$. Within this topological phase   MBSs appear at the ends of the TI nanowire, due to both the large subband-splitting energy $E_{\rm so}$ and ease of opening a field induced gap $\delta(\varphi)$ the phase space for MBSs can be exceptionally large ($\sim 20$~meV).

The key features of TI nanowires that enable a large topological superconducting phase are the large splitting energy due to a non-uniform potential $E_{\rm so}\sim\varepsilon_0\sim20$~meV and the field-induced  large gap due to orbital effects $\delta(\varphi)\sim 3$~meV/T (equivalent to a g-factor $g\sim 100$). However, in the 
limit of strong coupling between TI and superconductor 
%strong coupling limit 
it can be expected that metallization effects on the TI nanowire due to the  superconductor will modify these desirable properties. In the remainder of this manuscript we will investigate the impact of these metallization effects in the strong coupling regime  and the challenge they pose to achieving a topological superconducting phase in the TI nanowire.

\section{Model} \label{sec:model}
To model the superconducting proximity effect and metallization effects in the setup shown in Fig.~\ref{fig:setup}, we use a tight binding model with translational invariance along the length of the system ($y$-direction in Fig.~\ref{fig:setup}) such that the momentum $k$ parallel to the nanowire remains a good quantum number \cite{reeg2018}. The total Hamiltonian of the coupled superconductor and TI nanowire can therefore be written as $H=\sum_k H_k$, where for each $k$ we define 
\begin{equation}
H_k=H^{\rm TI}_k+H^{\rm SC}_k+H^{c}_k,
\end{equation}
where $H^{\rm TI}_k$ is the Hamiltonian of the TI nanowire, $H^{\rm SC}_k$ is the Hamiltonian of the superconductor, and $H^{\rm c}_k$ describes the tunnel coupling between TI nanowire and superconductor.

\begin{figure}[!t]
	\centering
  \includegraphics[width=1\columnwidth]{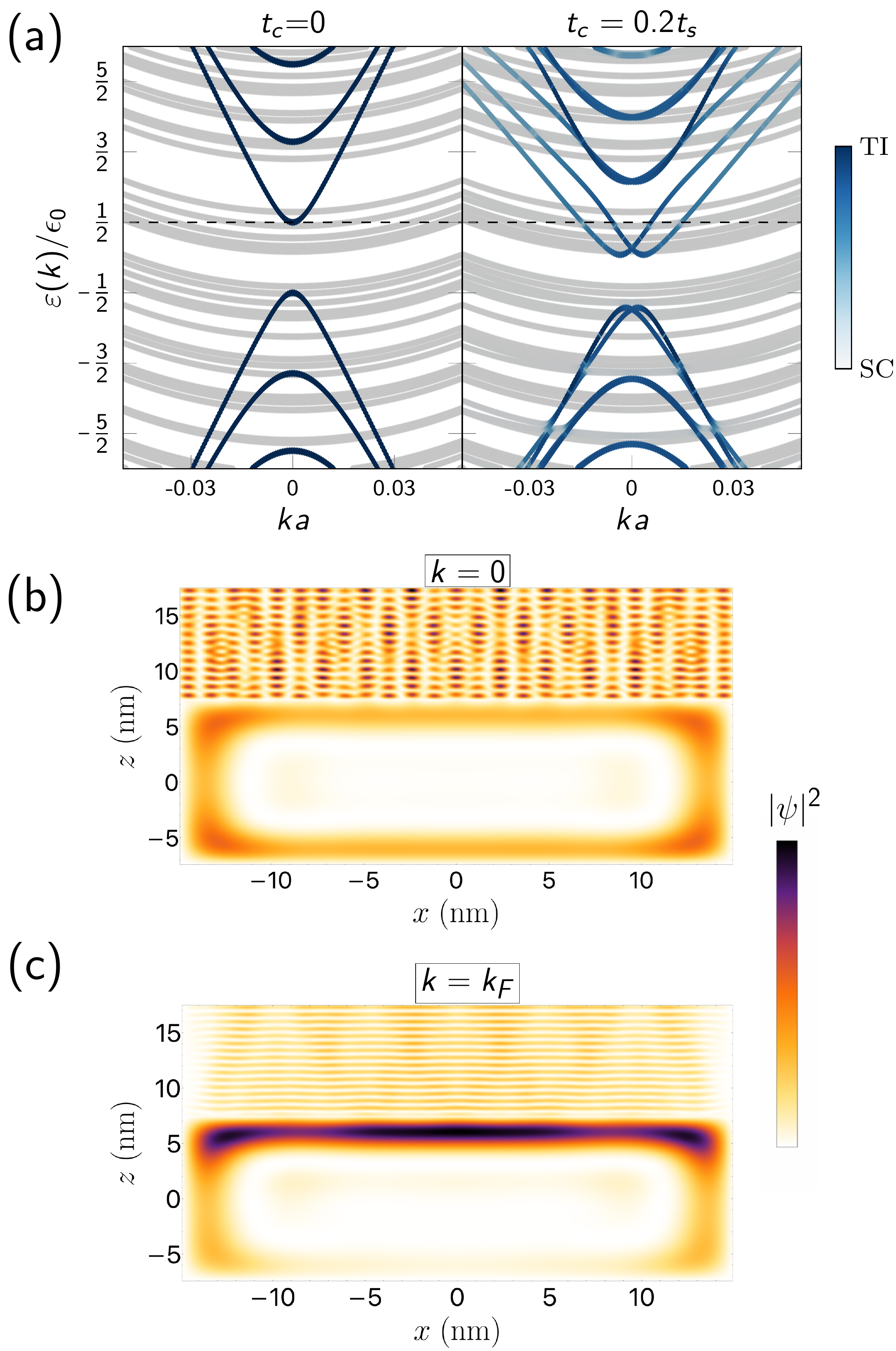}
	\caption{{\bf Tunnel coupling a TI nanowire to a superconductor:} Setup: Superconductor only on top of the TI nanowire [Fig.~\ref{fig:setup}(a)] as well as we set the pairing potential to zero, $\Delta_0=0$. {\bf (a)} Left: Subbands of TI nanowire (blue lines) and of superconducting layer (gray bands) with no tunnel coupling. The subbands of the TI are doubly degenerate and equally spaced (for our parameters $\epsilon_0\approx22$~meV). Right: A finite tunnel coupling induces an effective shift in the chemical potential close to the top of the nanowire and results in a shift of the TI nanowire subbands and, in addition, the non-uniformity of the potential causes a splitting of these subbands at finite momentum. The subbands of the superconductor remain largely unchanged. {\bf (b)} The probability density $|\psi|^2$ of the state corresponding to the subband crossing at $k=0$ for a finite coupling $t_c=0.2t_s$. The color code indicates the square modulus of the wavefunction. Despite the tunnel coupling to the superconductor, the state in the TI is equally localized at the top and bottom of the nanowire and a portion of the wavefunction is located in the superconductor -- this portion is  distinguishable by the small wavelength oscillations in the superconductor.  {\bf (c)} The same as in the panel (b) but for the state at finite Fermi momentum $k_F$ where the chemical potential $\mu_0$ is kept at the subband-crossing point. This finite momentum state is localized to top of the nanowire due to the potential induced by the superconductor, similar to the effect of a back-gate. Remaining model parameters same for all plots (see Sec.~\ref{sec:model}).}
	\label{fig:bands}
\end{figure}

To model the TI we use the BHZ Hamiltonian \cite{bernevig2006,zhang2009,Liu2010} on a square lattice with a rectangular cross-section, such that
\begin{align}
&H^{\rm TI}_k=\frac{1}{2}\sum_{\substack{n=1\\m=1}}^{N,M}{\bf c}^\dagger_{n,m,k}\!\cdot\!\{M(k)\tau_z+\frac{A}{a} \sin(k a) \tau_x \sigma_x  \}\,{\bf c}_{n,m,k} \nonumber \\
&+\sum_{\substack{n=1\\m=1}}^{N-1,M} \!\left\{{\bf c}^\dagger_{n+1,m,k}\!\cdot\! \left\{\frac{B}{a^2}\tau_x + \frac{i A}{2 a} \tau_x   \sigma_z \right\}e^{i  \phi^x_{m} }  {\bf c}_{n,m,k}\right\} \nonumber \\
&+\sum_{\substack{n=1\\m=1}}^{N,M-1} \!\left\{{\bf c}^\dagger_{n,m+1,k}\!\cdot\! \left\{\frac{B}{a^2}\tau_x + \frac{i A}{2 a} \tau_y \right\}e^{i \phi^z_{n} } {\bf c}_{n,m,k}\right\}\nonumber\\
&-\frac{1}{2}\sum_{\substack{n=1\\m=1}}^{N,M} \mu_{n,m} {\bf c}^\dagger_{n,m,k}\cdot {\bf c}_{n,m,k} \;\;+\text{H.c.},\label{seq:BHZ}
\end{align}
where $M(k)=M_0-2B[\cos(k a)-3]/a^2$. Here,  ${\bf c}^\dagger_{n,m,k}=(c^\dagger_{+,\uparrow},c^\dagger_{-,\uparrow},c^\dagger_{+,\downarrow},c^\dagger_{-,\downarrow})_{n,m,k}$ is a 4-vector, where $c^\dagger_{+(-),\uparrow(\downarrow)}$ describes the creation of an electron~$+$ (hole~$-$) with spin $\uparrow$ $(\downarrow)$ on site $(n,m)$ in the cross-section and with momentum $k$ along the nanowire. The influence of a gate will be modelled by the local potential $\mu_{n,m}$. Throughout we choose the lattice constant $a=1.5$~\AA~  such that the rectangular cross-section is  of width $w=(N-1) a$ and height $h=(M-1) a$ with $M=100$ and $N=200$, which means that the nanowire cross-section is approximately of dimension $30\;{\rm nm}\times 15\;{\rm nm}$. Orbital effects due to the magnetic field parallel to the nanowire, which open a gap at $k=0$, are taken into account through the Peierls phases $ \phi^{x}_{m}=\frac{a\pi}{hw}\left[aM-(h+1)/2\right]$ and $ \phi^{z}_{n}=\frac{a\pi}{hw}\left[aN-(w+1)/2\right]$. For clarity and simplicity we chose BHZ model parameters that are isotropic along all three axes, but such that they are consistent with Bi$_2$Se$_3$\cite{Liu2010}, namely $A=3.3$~meV\AA,  $B=44.5$~meV\AA$^2$, and $M_0=-280$~meV. Note that the finite extent of the wavefunction into the bulk of the nanowire means that the effective perimeter $P$ of the TI nanowire spectrum, Eq.~\eqref{eq:TIspec}, is smaller than this. As such we define the subband spacing $\epsilon_0$ using the gap between the subbands closest to the charge neutrality point of the BHZ model (see left panel of Fig.~\ref{fig:bands}).

The Hamiltonian of the superconductor is given by \cite{reeg2018} 

\begin{align}
&H^{\rm SC}_k=\sum_{n,m,\sigma} \big\{ 2 t_s\left[3\!-\!\cos(k a)\right]\!-\!\mu_s\big\} b^\dagger_{n,m,k,\sigma} b_{n,m,k,\sigma}\nonumber\\
&-t_s\left[\sum_{\substack{<n,n'>\\m,\sigma}}\!b^\dagger_{n',m,k,\sigma} b_{n,m,k,\sigma}+\sum_{\substack{<m',m>\\n,\sigma}} \!b^\dagger_{n,m',k,\sigma} b_{n,m,k,\sigma}\nonumber\right]\\
&+\sum_{n,m}(\Delta_0  b^\dagger_{n,m,k,\uparrow} b^\dagger_{n,m,-k,\downarrow}\!+\!\Delta_0^* b_{n,m,-k,\downarrow} b_{n,m,k,\uparrow}),\label{eq:hamsc}\end{align}
where $b^\dagger_{n,m,k,\sigma}$ and $b_{n,m,k,\sigma}$ are the corresponding creation and annihilation operators for the electrons in the superconductor, respectively,
$\sigma=\uparrow,\downarrow$ the spin values of the electrons along the quantization axis,
%are the spin degree of freedom 
and where we choose the lattice constant of the superconductor, $a=1.5$~\AA, to be the same as that used in the BHZ   model for the TI (see above). We consider different shapes and heights of superconductors, therefore the sums run over all sites $(n,m)$ of the superconductor and $\langle n,n' \rangle$ indicates a sum over nearest neighbour sites. For numerical simplicity we do not model any specific superconductor, however, throughout we use  parameters which give reasonable electronic properties compared to those found in, for example, niobium or vanadium. Namely, we choose the hopping $t_s=4$~eV and set the chemical potential in the superconductor to $\mu_s=0.5 t_s$, this corresponds to an effective mass $m^*_s=\hbar^2/(2t_s a^2)\approx 0.43\:m_e$ and Fermi velocity $v^{\rm sc}_F=\sqrt{2\mu/m^*}\approx 1.8\times10^6$~m/s.  

Finally, the tunnelling between the TI and superconductor is described by hopping from sites on either side of the interface between the two subsystems, these sites are denoted by $\langle i , j \rangle$. The coupling Hamiltonian between TI and superconductor is therefore given by 
 \begin{align}
 H^c_k = &- t_c\sum_{\substack{n,\sigma,\eta\\<m',m>}} b^\dagger_{n,m',k,\sigma} c_{n,m,k,\eta,\sigma}\\
 &- t_c \sum_{\substack{m,\sigma,\eta\\<n',n>}} b^\dagger_{n',m,k,\sigma} c_{n,m,k,\eta,\sigma}\label{eq:hamcoup}+{\rm H.c.},\nonumber
 \end{align}
where 
%$\sigma=\uparrow\downarrow$ is the spin and 
$\eta=\pm$ denotes the particle/hole degree of freedom of the BHZ model. As such, the hopping between adjacent sites preserves spin but otherwise the coupling $t_c$  is isotropic between the degrees of freedom of the TI and those of the superconductor.

\section{Metallization effects}\label{sec:metal}

To begin we consider a setup with the superconductor only on top of the TI nanowire as shown in Fig.~\ref{fig:setup}(a). We will discuss the potential advantages of the horseshoe-shaped superconductor shown in Fig.~\ref{fig:setup}(b) in the next section. 

\begin{figure*}[t]
	\centering
  \includegraphics[width=1.75\columnwidth]{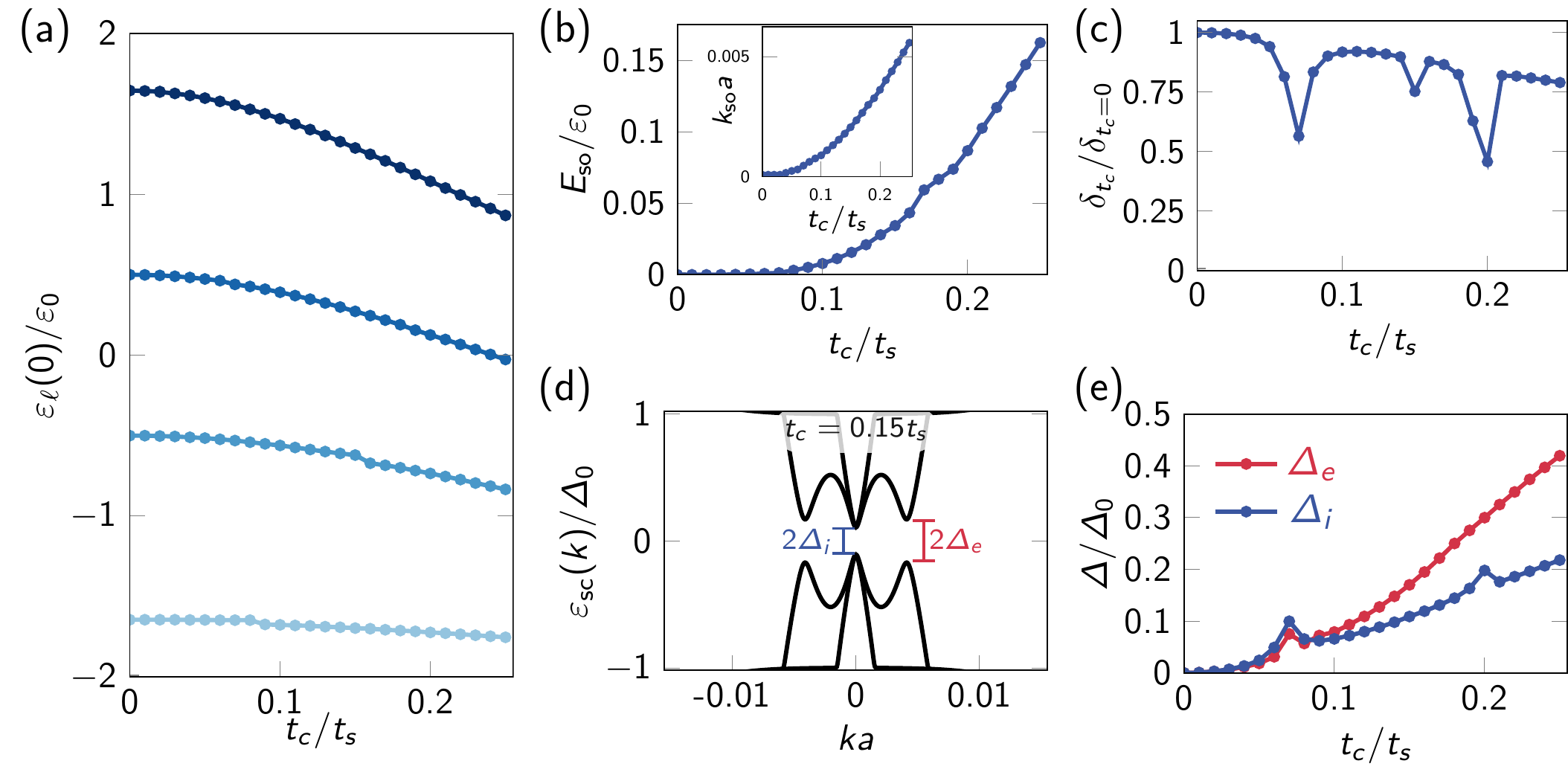}
	\caption{{\bf Metallization and proximity superconductivity in a TI nanowire:} Setup: Superconductor is placed only on top of the TI nanowire [see Fig.~\ref{fig:setup}(a)] and pairing potential $\Delta_0=1.5$~meV. {\bf (a)} The energy $\varepsilon_\ell(k=0)$ for the four subbands closest to the Dirac point as a function of tunnel coupling $t_c$. We observe a subband-shift up to $\sim 20$~meV, which is dependent on the subband index $\ell$. {\bf (b)} Evolution of the superconductor induced subband-splitting as a function of tunnel coupling $t_c$. This is parameterized by the splitting energy $E_{\rm so}$ and splitting momentum $k_{\rm so}$ (see Fig.~\ref{fig:schematic}). In the very strong coupling regime ($t_c/t_s\sim 0.25$) we find $E_{\rm so}\sim 4$~meV and splitting lengths $l_{\rm so}\equiv 1/k_{\rm so}\sim 25~$nm. {\bf (c)} Dependence of magnetic-field induced gap $\delta$ on tunnel coupling $t_c$. The magnetic-field induced gap $\delta(\varphi)$ is calculated using the gap at $k=0$ in the presence of a small flux $\varphi=10^{-4}$. The overall behavior shows a small reduction in the size of $\delta(\varphi)$ with larger reductions occurring when the TI states at $k=0$ are very close in energy to a subband of the superconductor and as a result strongly hybridize with it (for example at $t_c/t_s=0.2$, see also similar features in the panel (e). {\bf (d)} Spectrum of the the TI nanowire with the chemical potential tuned to the subband crossing at $k=0$ for a finite tunnel coupling $t_c=0.15t_s$. As outlined in Sec.~\ref{sec:tsc}, the interior and exterior proximity-induced gap, $\Delta_i$ and $\Delta_e$, respectively, stem from the pairing potential $\Delta_0$ inside the superconductor. These gaps are dependent on momentum due to the fact that the wavefunctions corresponding to large momenta are more strongly localized close to the top surface of the TI nanowire  (see Fig.~\ref{fig:bands}). {\bf (e)} Evolution of the interior (blue) and exterior (red) gap as a function of tunnel coupling $t_c$. The sizes of $\Delta_i$ and $\Delta_e$ are considerably different in the strong coupling regime where the subband-splitting due to the coupling with the superconductor is largest. Remaining model parameters same for all plots (see Sec.~\ref{sec:model}).}
	\label{fig:metal}
\end{figure*}

\subsection{Effect of tunnel coupling to superconductor}

We first consider the impact of coupling the superconductor to the TI nanowire in the absence of non-uniform potential generated by an applied gate voltage. The TI nanowire energy spectrum for a finite coupling $t_c$ is shown in Fig.~\ref{fig:bands}, the most notable features are a subband-shift similar in size to the subband spacing $\epsilon_0$, a subband-splitting at finite momenta, and a renormalization of Fermi velocity.

As shown in Fig.~\ref{fig:bands}, there is subband-shift when the superconducting layer is coupled to the TI nanowire, similar to the subband-shift previously investigated in semiconductors. This subband-shift is most easily quantified by considering the change of the subband energy at zero momentum, $\varepsilon_\ell(k=0)$. The full dependence of this shift in $\varepsilon_\ell(k=0)$ with respect to the tunnel coupling $t_c$ is shown in Fig.~\ref{fig:metal}(a). We find that this shift reaches values of the order of the subband spacing $\epsilon_0 \sim 20$~meV for realistic parameters, which is smaller than the effect in semiconductors \cite{reeg2018}. We also observe that the shift of the subbands is always $n$-type and is strongly dependent on the subband index $\ell>0$.

Possibly the most striking feature of Fig.~\ref{fig:bands} is the subband-splitting resulting from the coupling to the superconductor. As discussed in Sec.~\ref{sec:tsc}, a non-uniform potential through the nanowire cross-section can induce such a subband-splitting. Above we considered a non-uniformity created due to gating, however, the potential resulting from a superconductor attached to only one side of the nanowire also results in a sizable non-uniform potential with the effect largest close to the interface. As a result, the coupling to a superconductor by itself results in a subband splitting of size $E_{\rm so}$. The evolution of the splitting-energy as a function of increasing coupling strength $t_c$ is shown in Fig.~\ref{fig:metal}(b). Additionally, since for small momenta only the average of the potential induced by the superconductor, $\mu_0$, results in a subband-shift, the large non-uniformity of the potential also partially explains why the overall subband-shift (discussed above) is smaller compared to that found in semiconductors.

The size of the field-induced gap at $k=0$, $\delta(\varphi)$, is also modified by the tunnel coupling to a superconductor. As shown in Fig.~\ref{fig:metal}(c) it is renormalized down and substantial reductions occur when the crossing point at $k=0$ coincides with the bottom of subband of the superconductor [resulting in the dips shown in Fig.~\ref{fig:metal}(c)]. We find that the size of the gap is directly proportional to the weight of the TI state in the superconductor.

Moving now to the discussion of the size of the proximity-induced pairing potential $\Delta$ in the TI nanowire, in Fig. \ref{fig:metal}(d), we analyze  the typical spectra for the chemical potential tuned to the subband crossing point at $k=0$. We see that the size of the resulting superconducting gap is $k$-dependent leading to an interior gap $\Delta_i$ close to $k=0$ to be of different size compared to the exterior gap $\Delta_e$. The reason for the momentum-dependent gap stems from the fact that finite momenta states are more localized at the top of the TI compared to the states near $k=0$ (see Fig.~\ref{fig:metal}) and therefore states at large finite momenta couple more efficiently to the superconductor (note that if a gate voltage is applied the opposite effect can occur, see below). When the chemical potential is tuned close to the subband crossing point of the lowest subband, this momentum dependence can reduce the magnetic field strength required to achieve the topological superconducting phase, in accordance with the topological phase transition criterion defined in Eq.~\eqref{eq:topcrit}. This is a useful effect as it helps to reduce the detrimental impact of the magnetic field on the parent superconductor itself.

\subsection{Impact of gating} 
\begin{figure}[t]
	\centering
  \includegraphics[width=1\columnwidth]{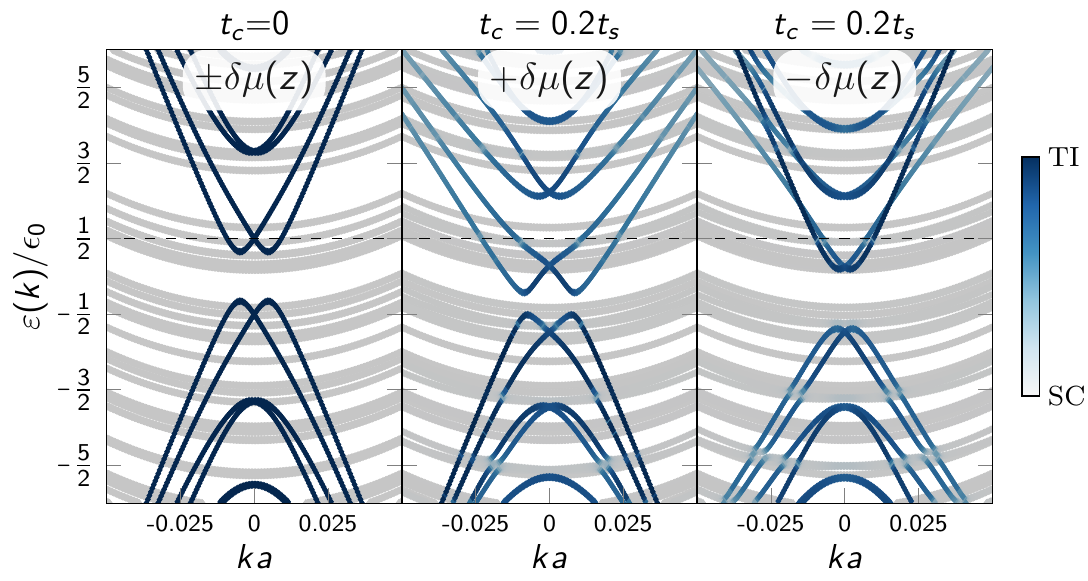}
	\caption{
	{\bf Tunnel coupling of a gated TI nanowire to a superconductor:} Setup: Superconductor only on top of the TI nanowire [Fig.~\ref{fig:setup}(a)] as well as we set the pairing potential to zero, $\Delta_0=0$. Left: gating the nanowire induces an non-uniform chemical potential through the cross-section, modelled here by $\pm\delta\mu(z)=\mp 2 \epsilon_0z/h$. Center: If the potential induced on the top of the nanowire by a finite tunnel coupling between TI and superconductor has the same sign as that induced by the gate then the subband-splitting is enhanced by the presence of the superconductor. Right: If the potentials induce by the gate and by the superconductor compete then the subband-splitting is reduced. Remaining model parameters same for all plots (see Sec.~\ref{sec:model}).}
	\label{fig:gated}
\end{figure}

We now consider the interplay of gating and metallization effects. As discussed in the Sec.~\ref{sec:tsc}, the application of a gate breaks inversion symmetry which lifts the degeneracy of subbands at finite momenta and results in a splitting of the subbands with individual states localized close to or far from the gate, depending on the sign of the applied gate voltage (see Fig.~\ref{fig:gated} far left panel). 

\begin{figure}[t]
	\centering
  \includegraphics[width=1\columnwidth]{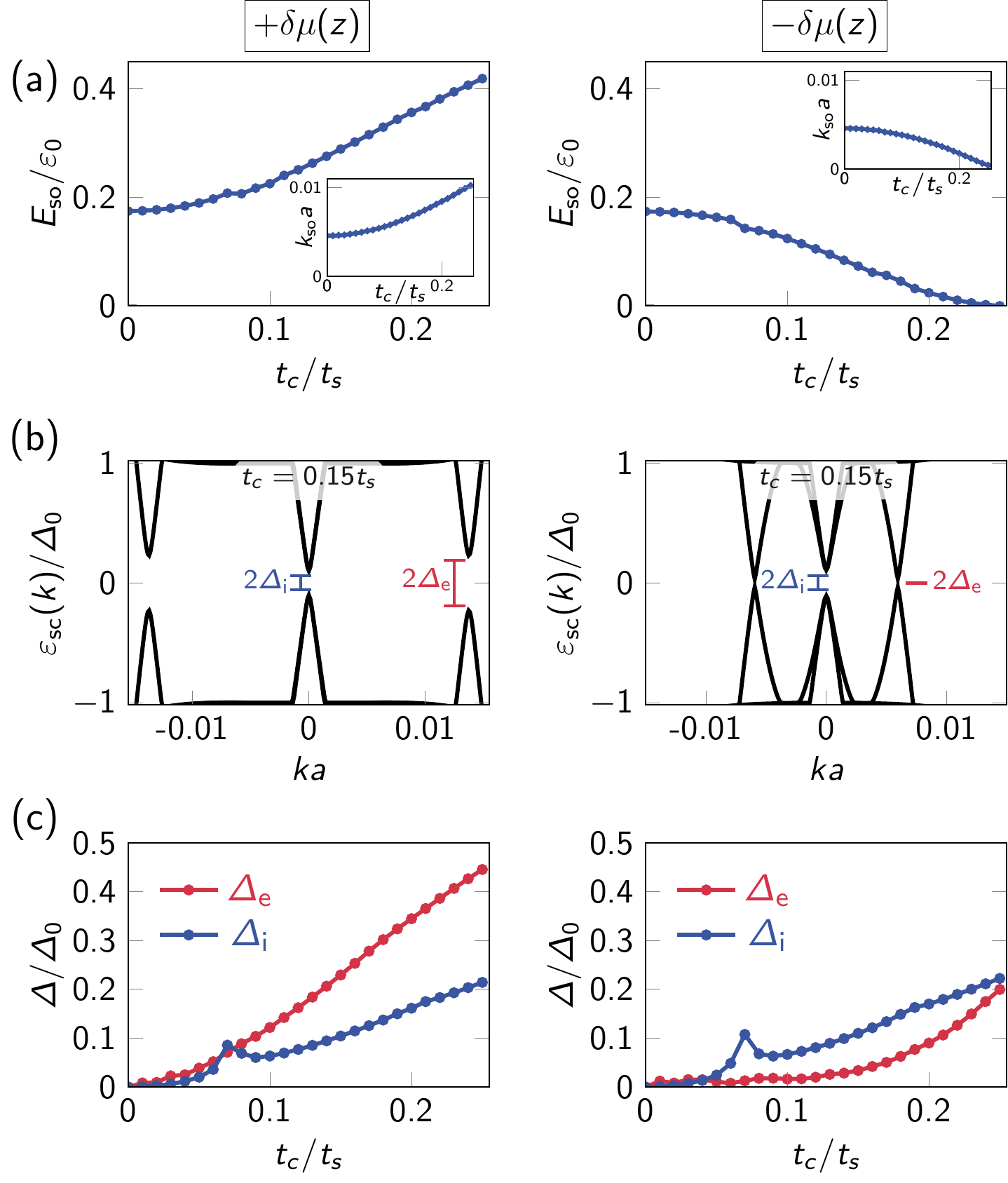}
	\caption{{\bf Impact of gating on metallization and proximity superconductivity in a TI nanowire:} Setup: Superconductor only on top of the TI nanowire [Fig.~\ref{fig:setup}(a)] and pairing potential is fixed to $\Delta_0=1.5$~meV. Left column: Impact of a gate induced non-uniform potential modelled by $+\delta\mu(z)=- 2 \epsilon_0z/h$. Right column: Impact of a gate induced non-uniform potential modelled by $-\delta\mu(z)=+ 2 \epsilon_0z/h$. {\bf (a)}~Evolution of the subband-splitting energy, $E_{\rm so}$, and splitting momentum, $k_{\rm so}$, as a function of tunnel coupling $t_c$. Depending on the relative sign of the gate induced potential and the potential arising from the coupling to the superconductor, the subband-splitting can either be enhanced or diminished (see Fig.~\ref{fig:gated}). {\bf (b)} Spectra of the the TI nanowire with the chemical potential tuned to the subband crossing at $k=0$ for a finite tunnel coupling. The interior gap $\Delta_i$ is similar for both signs of gate induced potential, however, the exterior gap $\Delta_e$ depends strongly on the gate potential. In particular we find a negligibly small exterior gap when the gate and superconductor induced non-uniform potentials compete (right column). {\bf (c)} Evolution of the interior (blue) and exterior (red) gaps [defined in panel (b)] as a function of tunnel coupling. When the superconductor- and gate-induced potentials compete, a sizeable exterior gap is only appears in the spectrum at the values of $t_c$ at which $E_{\rm so}\sim 0$. Remaining model parameters same for all plots (see Sec.~\ref{sec:model}).}
	\label{fig:renormalizationsgated}
\end{figure}

As discussed above, the induced potential at the interface with the superconductor can result in a similar localization of the states on one surface of the TI nanowire and associated splitting of the TI nanowire subbands. To model this we use a potential which varies in the vertical ($z$-direction) through the TI nanowire cross-section such that $\pm\delta\mu(z)=\mp 2 \epsilon_0z/h$ with $z$ measured from the center of the nanowire. It was previously shown that the induced subband splitting is not strongly dependent on the exact shape of the gate induced non-uniform potential $\delta\mu(z)$ (see Ref.~\cite{Legg2021}).  We find that the interplay between the gate voltage and potential arising from the superconductor can either enhance or reduce the size of the subband splitting, $E_{\rm so}$, depending on the relative sign of the induced potentials from the gate and superconductor, see Fig.~\ref{fig:gated}. If the sign of the non-uniform potential induced by the gate and by superconductor on the surface adjacent the superconductor are opposite the total subband-splitting is reduced, however, if the signs are the same the subband-splitting is enhanced. We note that this is in strong contrast with semiconductors, where the equivalent subband-splitting due to spin-orbit coupling is a materials property and is therefore reduced by metallization effects \cite{reeg2018}. A plot of this splitting energy as a function of the tunnel coupling strength $t_c$ is shown in Fig.~\ref{fig:renormalizationsgated}(a) for both signs of gate induced potential. 

The impact the non-uniform potential, $\delta\mu(z)$, arising due to gating, on the induced pairing potential, $\Delta$, is shown in Fig.~\ref{fig:renormalizationsgated}(b). Here, the chemical potential is kept at the subband crossing point at $k=0$ of the lowest subband. Whilst the interior gap $\Delta_i$ is similar in magnitude for both signs of gate potential, the exterior gap $\Delta_e$ differs dramatically. Strikingly, a negligibly small exterior gap is present when the gate and superconductor-induced potentials compete [$-\delta\mu(z)$], whereas when the potentials complement each other [$+\delta\mu(z)$] the exterior gap is much larger than the interior gap in the strong coupling regime. Finally, in Fig.~\ref{fig:renormalizationsgated}(c), we demonstrate the evolution of the interior and exterior gaps as a function of the tunnel coupling $t_c$. When the gate-induced and superconductor-induced non-uniform potentials complement each other, the evolution of the induced gap $\Delta$ is similar to that of the ungated nanowire [left panel of Fig.~\ref{fig:renormalizationsgated}(c)], whilst when the potentials compete the exterior gap $\Delta_e$ is negligible up to the tunnelling strengths at which $E_{\rm so} \approx 0$ [right panel Fig.~\ref{fig:renormalizationsgated}(c)].

\begin{figure}[t]
\centering
  \includegraphics[width=0.95\columnwidth]{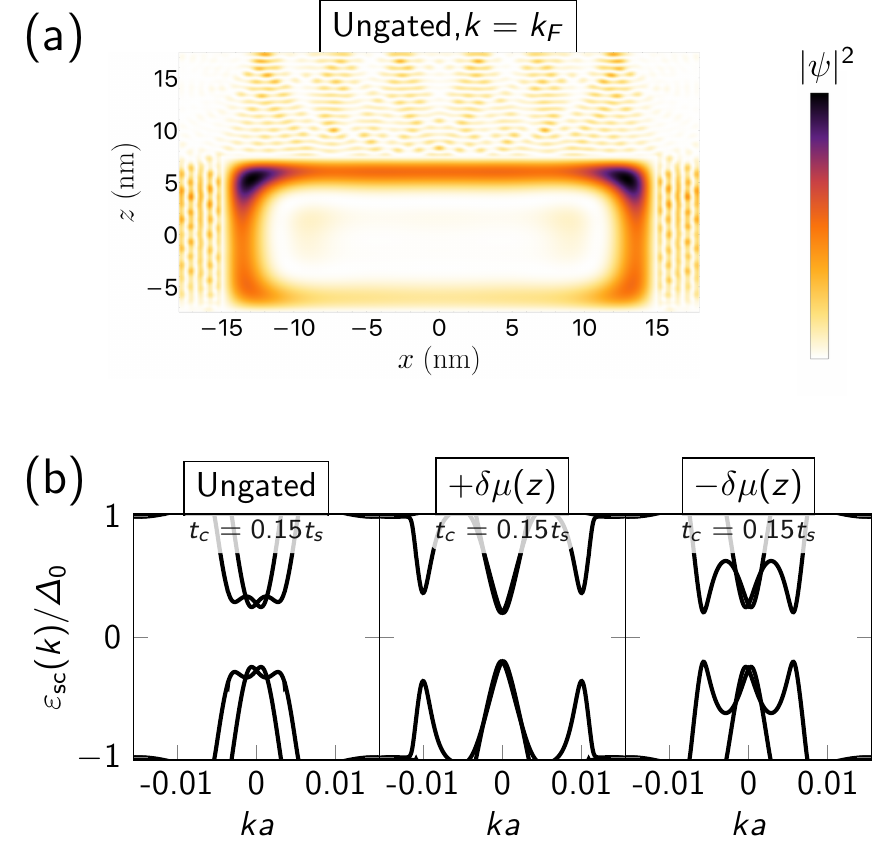}
	\caption{{\bf Potential advantages of a horseshoe superconductor:} Superconductor covers multiple surfaces of the TI nanowire [see Fig.~\ref{fig:setup}(b)] with a thin side layer of thickness of 3~nm and $\Delta_0=1.5$~meV. {\bf (a)} Typical probability density profile of the wavefunction of a TI state at finite momentum for tunnel coupling $t_c =0.15 t_s$. The wavefunction is localized on all surfaces of the TI nanowire that are connected to the superconductor.  {\bf (b)} Spectra of the TI nanowire with the chemical potential tuned to the subband crossing at $k=0$ for a finite tunnel coupling $t_c$. The size of the interior gap $\Delta_i$ stays almost the same independent of gating. In addition, a sizable exterior gap is induced regardless whether a gate voltage is applied or not as well as  the sign of the applied gate voltage does not affect the strength of the superconducting proximity effect. In contrast to that, the size of the subband splitting strongly depends on the gating. Remaining model parameters same for all plots (see Sec.~\ref{sec:model}).}
	\label{fig:horseshoe}
\end{figure}

\section{Experimental considerations and comparison with semiconductors}\label{sec:exp}

\subsection{Metallization effects}
In the previous section we saw that a strong tunnel coupling between a superconducting layer and a TI nanowire leads to a shift and splitting of the TI nanowire subbands as well as a reduction of the size of the possible gap $\delta(\varphi)$ opened by orbital effects when a magnetic field is applied parallel to the nanowire. In this section we will consider the challenges posed by and possible mitigations of these effects possible in experiments. The main criteria for achieving MBSs in TI nanowires are: 1)~A large exterior gap $\Delta_e \sim\Delta_0$. 2)~The ability to tune the chemical potential close to the subband crossing.  3)~A large splitting energy $E_{\rm so}$ such that helical modes cross the Fermi-level at large finite momentum. 4)~A magnetic field induced gap $\delta(\varphi)$ which can exceed the size of the interior superconducting gap $\Delta_i$ at field strengths lower than the critical field of the superconductor, such that the topological criterion in Eq.~\eqref{eq:topcrit} can be satisfied. 

We first note that our results show that, unlike in semiconductor nanowires, metallization effects in TI nanowires pose a more limited experimental challenge to the ability to fulfil the requirements for topological superconductivity. For instance, in the strong coupling limit $\Delta\sim\Delta_0$, the subband-shift ($\sim 20$~meV) found in TI nanowires is smaller than that found in semiconductors ($\sim 100$~meV). A shift of this size can relatively easily be mitigated by an applied gate voltage \cite{Munning2021} or a change of the compensation of the TI material during growth \cite{ren2011,zhang2011}.

Further, as long as one is careful about the sign of the applied gate voltage or uses a horseshoe geometry (see below), in the strong coupling limit a large splitting energy $E_{\rm so}$ is maintained and even enhanced by metallization effects in TI nanowires. The main reason stems from the fact that the splitting energy in TI nanowires is induced externally and is not a material property. This presents a significant benefit of a TI nanowire compared to a semiconductor with strong Rashba spin orbit coupling since, unlike in TI nanowires, the subband splitting due to spin-orbit coupling in a semiconductor will always be renormalized by the smaller spin-orbit effects of common superconductors. The tunability of the splitting energy in TI nanowires also makes it possible to further mitigate the effects of the superconductor by simply inducing a large or smaller non-uniform potential,  this is not possible in semiconductor nanowires. If required, this additional tunability could be achieved by, for example, the addition of a top gate \cite{Yang2015,taskin2017} which would provide a more independent tuning of the average chemical potential $\mu_0$ and the non-uniform potential $\delta\mu(\theta)$. Note that such dual gate device could also make it easier to ensure a large exterior gap (see below).

Finally, the reduction of the magnetic field induced gap $\delta(\varphi)$ is also a relatively minor problem in TI nanowires. This is because $\delta(\varphi)$ is primarily created by orbital effects and therefore the size of the effective $g$-factor (before metallization) is huge, for example, the $30\;{\rm nm}\times 15\;{\rm nm}$ TI nanowires considered here have effective $g$-factor $\sim 100$. Further, if desired, the reduction of $\delta(\varphi)$ can be offset relatively easily simply by utilising a nanowire with a larger cross-section. This is in stark contrast to semiconductor nanowires where the $g$-factor (in the Zeeman term) has a very limited tunability if one wants to offset metallization effects.

\subsection{Geometry of superconductor}
A significantly smaller interior gap $\Delta_i$ than exterior gap $\Delta_e$ -- as found when the non-uniform potential from the superconductor and from the gate complement each other -- is preferable since it reduces the field required to reach the topological phase as defined in Eq.~\eqref{eq:topcrit}. Nonetheless, in the previous section, we also found that inducing an exterior gap $\Delta_e\sim\Delta_0$ at all momenta is not always possible, see Fig.~\ref{fig:renormalizationsgated}. The main reason for this is 
%because 
that a large superconducting pairing term is only strongly induced for states in the TI nanowire  that have a sizable probability density adjacent to the superconductor. This presents a limitation to the versatility of devices with a superconductor on only one side of the TI nanowire [as shown in Fig.~\ref{fig:setup}(a)] since the applied gate voltage must always induce a non-uniform potential that for large momenta localizes the states  to the top of the nanowire cross-section, this is only easily achievable for all chemical potentials in a dual gate device \cite{Yang2015,taskin2017}. 

If one wants to ensure that a sizable exterior gap $\Delta_e$ is always present, regardless of the non-uniform potential, it is better to attach the superconductor to more than one side of the TI nanowire. To highlight this, we have also considered a horseshoe superconductor geometry [shown in Fig.~\ref{fig:setup}(b)] where the superconductor is attached to the top and side surfaces of our rectangular TI nanowire cross-section. The potential advantages of a horseshoe geometry are shown in Fig.~\ref{fig:horseshoe}. Namely, as in Fig.~\ref{fig:horseshoe}(a), we see that since the superconductor now couples to the sides of the nanowire and the TI wavefunction at finite momentum is now localized on all three sides of the nanowire cross-section. As a result, a sizable exterior gap $\Delta_e$ is induced regardless of the sign of the applied gate voltage due to the bottom gate [see Fig~\ref{fig:horseshoe}(b)], although it should be noted that a weak momentum dependence of the induced gap is still visible even in this setup. 

\begin{figure}[t]
\centering
 \hspace{-30pt} \includegraphics[width=0.9\columnwidth]{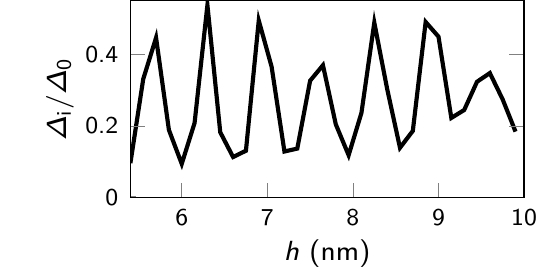}
	\caption{{\bf Dependence of interior gap $\mathbf{\Delta}_i$ on superconductor layer height $h$:} Setup: Superconductor only on top of the TI nanowire [Fig.~\ref{fig:setup}(a)], pairing potential is fixed to $\Delta_0=1.5$~meV, tunnel coupling $t_c=0.2t_s$, and no gating effects included. Changing the  finite height $h$ of the superconducting layer, we find that the size of proximity induced gap and metallization effects oscillate with a period set by the Fermi wavelength of the superconductor $\lambda_s$ (for our parameters $\lambda_s\approx0.9$~nm).
	Remaining model parameters same for all plots (see Sec.~\ref{sec:model}).}
	\label{fig:hdep}
\end{figure}

\subsection{Height dependence}
Another important effect arises from the finite height of the superconducting layer. Similar to semiconductor nanowires \cite{reeg2018}, the relative energy between the subbands within the superconducting layer and the TI nanowire subbands strongly impacts metallization and proximity effects. In particular, properties induced by the superconductor in the TI can be expected to oscillate as a function the layer height $h$. For instance this is shown in Fig.~\ref{fig:hdep} for the interior gap $\Delta_i$. Such oscillations occur with a period set by the Fermi wavelength of the superconductor $\lambda_s$ (for our parameters $\lambda_s\approx0.9$~nm). Although, due to disorder, one will observe some averaging of this effect, the oscillations of the TI nanowire properties due to small changes of layer thickness will likely make devices with consistent properties difficult to fabricate.

\section{Discussion} \label{sec:dis}
We have investigated the impact of tunnel coupling a superconducting layer to a TI nanowire. In the strong coupling regime, where the proximity induced pairing potential is of the order of that found in the parent superconductor, we found a distinct phenomenology associated with the metallization effects due to the tunnel coupling. Although we find a subband-shift similar in size to the TI nanowire subband gap $\sim 20$~meV and a reduction of the magnetic field induced gap $\delta(\varphi)$, these metallization effects are considerably smaller than the equivalents  in semiconductors and can be mitgated relatively easily by gating or compensation. We also found that, in certain scenarios, metallization effects can actually be beneficial for inducing a topological superconducting phase, for instance the subband-splitting associated with the breaking of inversion symmetry by an applied gate voltage can actually be enhanced by a similar potential arising at the interface between the TI nanowire and superconductor. 

Whilst metallization effects in TI nanowires do pose some experimental challenges we have shown that these can largely be mitigated during the device fabrication phase and do not present a significant roadblock to realizing MBSs. When compared to semiconductors, the reduced impact of a superconductor on the ability for a TI nanowire to realize topological superconductivity is a significant benefit, especially given recent developments in device fabrication and improvements of superconductor-TI interfaces \cite{Breunig2021}. 

The main remaining experimental challenge to achieve topological superconductivity and associated MBSs in TI nanowires is disorder. For instance it remains an open experimental question whether disorder, in particular, charged impurities in the bulk of TI nanowires \cite{skinner2013,skinner2013b,Borgwardt2016,Knispel2017,bomerich2017,huang2021}, will allow current TI nanowires to produce devices that can reliably and reproducibly realize MBSs. That said, the presence of a superconductor should help screen charged impurities and the large topological phase space for MBSs reduces constraints imposed by disorder.  Nonetheless, the reduced metallization effects studied here and the large scope for further improvements of TI materials makes TI nanowires a very competitive platform to realize MBSs in the near term.

%==============================================================================================
%    Acknowledgments
%==============================================================================================

\acknowledgments

We acknowledge useful discussions with Y. Ando, O. Breunig, and M. R\"o\ss ler. This work was supported by the Georg H. Endress Foundation and the Swiss National Science Foundation and NCCR QSIT. This project received funding from the European Union’s Horizon 2020 research and innovation program (ERC Starting Grant, Grant No 757725).

%==============================================================================================
%    Bibliography
%==============================================================================================

\bibliography{TI-nws}

%==============================================================================================
%    Supplemental Material (if any)
%==============================================================================================

%\input{supp.tex}

\end{document}